\documentclass[12pt]{article}
\usepackage{amsmath,amsthm}
\usepackage{eucal}
\usepackage{amsfonts}
\newcommand{\ar}{\renewcommand{\arraystretch}{1}} 
\gdef\C{\Bbb C}
\gdef\dZ{\Bbb Z}

\gdef\dS{\Bbb S}
\gdef\R{\Bbb R}

\gdef\F{\Bbb F}

\DeclareMathOperator{\Ker}{Ker}

\DeclareMathOperator{\Sym}{Sym}

\newcommand{\s}{\scriptstyle}
\newcommand{\iddots}{.\hspace{0.5mm}\raisebox{1mm}{.}\hspace{0.5mm}
\raisebox{2mm}{.}}
\newcommand{\tg}{\tan}
\newcommand{\ch}{\cosh}
\newcommand{\sh}{\sinh}
\newcommand{\tnh}{\tanh}

\newcommand{\re}{\mbox{\rm Re}\,}
\newcommand{\im}{\mbox{\rm Im}\,}

\newcommand{\cA}{\mathcal{A}}

\newcommand{\cE}{\mathcal{E}}

\newcommand{\M}{{\bf\sf M}}

\newcommand{\sA}{{\sf A}}
\newcommand{\sB}{{\sf B}}
\newcommand{\sI}{{\sf I}}

\newcommand{\sL}{\Lambda}

\newcommand{\sY}{{\sf Y}}
\newcommand{\sX}{{\sf X}}

\newcommand{\fM}{\mathfrak{M}}

\newcommand{\fG}{\mathfrak{G}}
\newcommand{\fT}{\mathfrak{M}}
\newcommand{\fg}{\mathfrak{g}}

\newcommand{\hypergeom}[5]{\mbox{$
_#1 F_#2 \left.
\!\!
\left(
\!\!\!\!
\begin{array}{c}
\multicolumn{1}{c}{\begin{array}{c}
#3
\end{array}}\\[1mm]
\multicolumn{1}{c}{\begin{array}{c}
#4
   \end{array}}\end{array}
\!\!\!\!
\right|\displaystyle{#5}\right)
$}
}
\newcommand{\cl}{C\kern -0.2em \ell}

\newcommand{\ld}{\left[}
\newcommand{\rd}{\right]}

\begin{document}
\title{Relativistic Wave Equations in the Helicity Basis}
\author{V.~V. Varlamov\thanks{Department of Mathematics, Siberia State
University of Industry, Kirova 42, Novokuznetsk 654007, Russia.}}
\date{}
\maketitle
\begin{abstract}
The principal series of unitary representations of the Lorentz group
has been considered in the helicity basis. Decompositions of tensor
products of the spinspaces are studied in the framework of projective
representations of the symmetric group. Higher-spin Gel'fand-Yaglom
equations are defined in the helicity basis over an arbitrary
representation space. Applications of decomposable and indecomposable
Gel'fand-Yaglom equations to particle physics are discussed.
\end{abstract}
\section{Introduction}
As known \cite{GG78}, a root subgroup of a semisimple Lie group $O_4$
(a rotation group of the 4-dimensional space) is a normal divisor of $O_4$.
For that reason the 6-parameter group $O_4$ is semisimple, and is
represented by a direct product of the two 3-parameter unimodular groups.
By analogy with the group $O_4$, a double covering $SL(2,\C)$ of the
proper orthochronous Lorentz group $\fG_+$ (a rotation group of the
4-dimensional Minkowski spacetime) is semisimple, and is represented by
a direct product of the two 3-parameter special unimodular groups,
$SL(2,\C)\simeq SU(2)\otimes SU(2)$. An explicit form of this isomorphism
can be obtained by means of a complexification of the group $SU(2)$,
that is, $SL(2,\C)\simeq\mbox{\sf complex}(SU(2))\simeq
SU(2)\otimes SU(2)$ \cite{Var022}.
Moreover, in the works \cite{AE93,Dvo96}
the Lorentz group is represented by a product $SU_R(2)\otimes SU_L(2)$,
and spinors
\[
\psi(p^\mu)=\ar\begin{pmatrix}
\phi_R(p^\mu)\\
\phi_L(p^\mu)
\end{pmatrix}
\]
are transformed within $(j,0)\oplus(0,j)$ representation space. The
components 
$\phi_R(p^\mu)$ and $\phi_L(p^\mu)$ correspond to different helicity
states (right- and left-handed spinors).

\section{Helicity basis}
Let $\mathfrak{g}\rightarrow T_{\mathfrak{g}}$ be an arbitrary linear
representation of group $\fG_+$ and let
$\sA_i(t)=T_{a_i(t)}$ be an infinitesimal operator corresponded the rotation
$a_i(t)\in\fG_+$. Analogously, we have $\sB_i(t)=T_{b_i(t)}$, where
$b_i(t)\in\fG_+$ is a hyperbolic rotation. The operators $\sA_i$ and
$\sB_i$ satisfy the following commutation 
relations:
\begin{equation}\label{Com1}
\left.\begin{array}{lll}
\ld\sA_1,\sA_2\rd=\sA_3, & \ld\sA_2,\sA_3\rd=\sA_1, &
\ld\sA_3,\sA_1\rd=\sA_2,\\[0.1cm]
\ld\sB_1,\sB_2\rd=-\sA_3, & \ld\sB_2,\sB_3\rd=-\sA_1, &
\ld\sB_3,\sB_1\rd=-\sA_2,\\[0.1cm]
\ld\sA_1,\sB_1\rd=0, & \ld\sA_2,\sB_2\rd=0, &
\ld\sA_3,\sB_3\rd=0,\\[0.1cm]
\ld\sA_1,\sB_2\rd=\sB_3, & \ld\sA_1,\sB_3\rd=-\sB_2, & \\[0.1cm]
\ld\sA_2,\sB_3\rd=\sB_1, & \ld\sA_2,\sB_1\rd=-\sB_3, & \\[0.1cm]
\ld\sA_3,\sB_1\rd=\sB_2, & \ld\sA_3,\sB_2\rd=-\sB_1. &
\end{array}\right\}
\end{equation}
Denoting $\sI^{23}=\sA_1$, $\sI^{31}=\sA_2$,
$\sI^{12}=\sA_3$, and $\sI^{01}=\sB_1$, $\sI^{02}=\sB_2$, $\sI^{03}=\sB_3$
we can write the relations (\ref{Com1}) in a more compact form:
\[
\ld\sI^{\mu\nu},\sI^{\lambda\rho}\rd=\delta_{\mu\rho}\sI^{\lambda\nu}+
\delta_{\nu\lambda}\sI^{\mu\rho}-\delta_{\nu\rho}\sI^{\mu\lambda}-
\delta_{\mu\lambda}\sI^{\nu\rho}.
\]

Let us consider the operators
\begin{gather}
\sX_k=\frac{1}{2}(\sA_k+i\sB_k),\quad\sY_k=\frac{1}{2}(\sA_k-i\sB_k),
\label{SL25}\\
(k=1,2,3).\nonumber
\end{gather}
Using the relations (\ref{Com1}) we find that
\begin{gather}
\ld\sX_1,\sX_2\rd=\sX_3,\quad\ld\sX_2,\sX_3\rd=\sX_1,\quad
\ld\sX_2,\sX_1\rd=\sX_2,\nonumber\\
\ld\sY_1,\sY_2\rd=\sY_3,\quad\ld\sY_2,\sY_3\rd=\sY_1,\quad
\ld\sY_3,\sY_1\rd=\sY_2,\nonumber\\
\ld\sX_k,\sY_l\rd=0,\quad(k,l=1,2,3).\label{Com2}
\end{gather}
Further, taking
\begin{equation}\label{SL26}
\left.\begin{array}{cc}
\sX_+=\sX_1+i\sX_2, & \sX_-=\sX_1-i\sX_2,\\[0.1cm]
\sY_+=\sY_1+i\sY_2, & \sY_-=\sY_1-i\sY_2
\end{array}\right\}
\end{equation}
we see that in virtue of commutativity of the relations (\ref{Com2}) a
space of an irreducible finite--dimensional representation of the group
$\fG_+$ can be stretched on the totality of $(2l+1)(2\dot{l}+1)$ basis
vectors $\mid l,m;\dot{l},\dot{m}\rangle$, where $l,m,\dot{l},\dot{m}$
are integer or half--integer numbers, $-l\leq m\leq l$,
$-\dot{l}\leq \dot{m}\leq \dot{l}$. Therefore,
\begin{eqnarray}
&&\sX_-\mid l,m;\dot{l},\dot{m}\rangle=
\sqrt{(\dot{l}+\dot{m})(\dot{l}-\dot{m}+1)}\mid l,m;\dot{l},\dot{m}-1
\rangle\;\;(\dot{m}>-\dot{l}),\nonumber\\
&&\sX_+\mid l,m;\dot{l},\dot{m}\rangle=
\sqrt{(\dot{l}-\dot{m})(\dot{l}+\dot{m}+1)}\mid l,m;\dot{l},\dot{m}+1
\rangle\;\;(\dot{m}<\dot{l}),\nonumber\\
&&\sX_3\mid l,m;\dot{l},\dot{m}\rangle=
\dot{m}\mid l,m;\dot{l},\dot{m}\rangle,\nonumber\\
&&\sY_-\mid l,m;\dot{l},\dot{m}\rangle=
\sqrt{(l+m)(l-m+1)}\mid l,m-1,\dot{l},\dot{m}\rangle
\;\;(m>-l),\nonumber\\
&&\sY_+\mid l,m;\dot{l},\dot{m}\rangle=
\sqrt{(l-m)(l+m+1)}\mid l,m+1;\dot{l},\dot{m}\rangle
\;\;(m<l),\nonumber\\
&&\sY_3\mid l,m;\dot{l},\dot{m}\rangle=
m\mid l,m;\dot{l},\dot{m}\rangle.\label{Waerden}
\end{eqnarray}
From the relations (\ref{Com2}) it follows that each of the sets of 
infinitisimal operators $\sX$ and $\sY$ generates the group $SU(2)$ and these
two groups commute with each other. Thus, from the relations (\ref{Com2})
and (\ref{Waerden}) it follows that the group $\fG_+$, in essence,
is equivalent to the group $SU(2)\otimes SU(2)$. In contrast to the
Gel'fand--Naimark representation for the Lorentz group \cite{GMS,Nai58}
which does not find a broad application in physics,
a representation (\ref{Waerden}) is a most useful in theoretical physics
(see, for example, \cite{AB,Sch61,RF,Ryd85}). This representation for the
Lorentz group was firstly given by Van der Waerden in his brilliant book
\cite{Wa32}.

As showed in \cite{Var021} the spinspace
\begin{equation}\label{Spin}
\dS_{2^{k+r}}\simeq
\underbrace{\dS_2\otimes\dS_2\otimes\cdots\otimes\dS_2}_{k\;\text{times}}
\otimes\underbrace{\dot{\dS}_2\otimes\dot{\dS}_2\otimes\cdots\dot{\dS}_2}_
{r\;\text{times}}
\end{equation}
is the minimal left ideal of the complex Clifford algebra
\begin{equation}\label{Cliff}
\C_{2n}\simeq
\underbrace{\C_2\otimes\C_2\otimes\cdots\otimes\C_2}_{k\;\text{times}}\otimes
\underbrace{\overset{\ast}{\C}_2\otimes\overset{\ast}{\C}_2\otimes\cdots
\otimes\overset{\ast}{\C}_2}_{r\;\text{times}}.
\end{equation}
The algebras $\C_2$ ($\overset{\ast}{\C}_2$) and the spinspaces
$\dS_2$ ($\dot{\dS}_2$) correspond to fundamental represesentations
$\boldsymbol{\tau}_{\frac{1}{2},0}$ ($\boldsymbol{\tau}_{0,\frac{1}{2}}$)
of the Lorentz group $\fG_+$.

In general case the spinspace (\ref{Spin}) is reducible, that is,
there exists a decomposition of the original spinspace $\dS_{2^{k+r}}$
into a direct sum of irreducible subspaces with respect to a
representation
\begin{equation}\label{Rep}
\underbrace{\boldsymbol{\tau}_{\frac{1}{2},0}\otimes
\boldsymbol{\tau}_{\frac{1}{2},0}\otimes\cdots\otimes
\boldsymbol{\tau}_{\frac{1}{2},0}}_{k\;\text{times}}\otimes
\underbrace{\boldsymbol{\tau}_{0,\frac{1}{2}}\otimes
\boldsymbol{\tau}_{0,\frac{1}{2}}\otimes\cdots\otimes
\boldsymbol{\tau}_{0,\frac{1}{2}}}_{r\;\text{times}}.
\end{equation}
First of all, let us consider a decomposition of the spinspace
\begin{equation}\label{Spin2}
\dS_{2^m}\simeq\dS_2\otimes\dS_2\otimes\cdots\otimes\dS_2
\end{equation}
into irreducible subspaces with respect to an action of the representation
\begin{equation}\label{Rep1}
\boldsymbol{\tau}_{\frac{1}{2},0}\otimes\boldsymbol{\tau}_{\frac{1}{2},0}
\otimes\cdots\otimes\boldsymbol{\tau}_{\frac{1}{2},0}.
\end{equation}
\subsection{Projective representations of the symmetric group}
A group ring $\dZ(S_m)$ and a group algebra $\C(S_m)$ 
of the symmetric group $S_m$
act in the spinspace (\ref{Spin2}). In 1911 \cite{Sch11} (see also
\cite{Mor62,Ste89}), Schur showed that over the field $\F=\C$ there exists
a nontrivial $\dZ_2$-extension $T_n$ of the symmetric group $S_n$ defined
by the sequence
\[
1\longrightarrow\dZ_2\longrightarrow T_n\longrightarrow S_n
\longrightarrow 1,
\]
where
\begin{multline}
T_n=\left\{z,\,t_1,\ldots,\,t_{n-1}:\;z^2=1,\; zt_i=t_iz,\right.\\
\left.t^2_i=z,\;\left(t_jt_{j+1}\right)^3=z,\; t_kt_l=zt_lt_k\right\}\\
1\leq i\leq n-1,\; 1\leq j\leq n-2,\; k\leq l-2.
\nonumber
\end{multline}
The group $T_n$ has order $2(n!)$. The subgroup $\dZ_2=\{1,z\}$ is central
and, is contained in the commutator subgroup of $T_n$,
$T_n/\dZ_2\simeq S_n$ ($n\geq 4$). If $n<4$, then every projective
representation of $S_n$ is projectively equivalent to a linear
representation.

Spinor representations of the transpositions $t_k$ are defined by elements
of the group algebra $\C(S_n)$:
\[
t_k=\sqrt{\frac{k-1}{2k}}\cE_{k-1}-\sqrt{\frac{k+1}{2k}}\cE_k,\quad
k=1,\ldots,m,
\]
where
\begin{equation}\label{6.6}
{\renewcommand{\arraystretch}{1.2}
\begin{array}{lcl}
\cE_{1}&=&\sigma_{1}\otimes\boldsymbol{1}_2\otimes\cdots\otimes
\boldsymbol{1}_2\otimes 
\boldsymbol{1}_2\otimes\boldsymbol{1}_2,\\
\cE_{2}&=&\sigma_{3}\otimes\sigma_{1}\otimes\boldsymbol{1}_2\otimes\cdots\otimes 
\boldsymbol{1}_2\otimes\boldsymbol{1}_2,\\
\cE_{3}&=&\sigma_{3}\otimes\sigma_{3}\otimes\sigma_{1}\otimes\boldsymbol{1}_2
\otimes\cdots\otimes\boldsymbol{1}_2,\\
\hdotsfor[2]{3}\\
\cE_{m}&=&\sigma_{3}\otimes\sigma_{3}\otimes\cdots\otimes\sigma_{3}
\otimes\sigma_{1},\\
\cE_{m+1}&=&\sigma_{2}\otimes\boldsymbol{1}_2\otimes\cdots\otimes\boldsymbol{1}_2,\\
\cE_{m+2}&=&\sigma_{3}\otimes\sigma_{2}\otimes\boldsymbol{1}_2\otimes\cdots\otimes 
\boldsymbol{1}_2,\\
\hdotsfor[2]{3}\\
\cE_{2m}&=&\sigma_{3}\otimes\sigma_{3}\otimes\cdots\otimes\sigma_{3}
\otimes\sigma_{2}.
\end{array}}\end{equation}
are the tensor poducts of $m$ Pauli matrices. These matrices form a spinor
basis of the complex Clifford algebra $\C_n$ ($n=2m$) in the
Brauer-Weyl representation \cite{BW35}. At this point there is an
isomorphism $\C_{2m}\simeq\M_{2^m}(\C)$, where $\M_{2^m}(\C)$ is a full
matrix algebra over the field $\F=\C$. If $n=2m+1$ we add one more matrix
\[
\cE_{2m+1}=\underbrace{\sigma_3\otimes\sigma_3\otimes\cdots\otimes
\sigma_3}_{m\;\text{times}}.
\]
In virtue of an isomorphism $\C_{2m+1}\simeq\C_{2m}\oplus\C_{2m}$
the units of $\C_{2m+1}$ are represented by direct sums
$\cE_1\oplus\cE_1$, $\cE_2\oplus\cE_2$, $\ldots$, $\cE_{2m}\oplus\cE_{2m}$.
Besides, there exists a mapping $\epsilon:\;\C_{2m+1}\rightarrow\C_{2m}$.
In the result of the homomorphism $\epsilon$ we have a quotient algebra
${}^\epsilon\C_{2m}\simeq\C_{2m+1}/\Ker\epsilon$, where
$\Ker\epsilon=\left\{\cA^1-\varepsilon\omega\cA^1\right\}$ is a kernel of 
$\epsilon$, $\cA^1$ is an arbitrary element of the subalgebra $\C_{2m}$,
$\omega$ is a volume element of $\C_{2m+1}$ \cite{Var021}.

Recently, projective representations of the symmetric group have been used
at the study of fractional quantum Hall effect \cite{NW96,Wil01,Wil02}
and non-Abelian statistics \cite{FG00}.

In accordance with a general Weyl scheme \cite{Weyl} a decomposition
of tensor products of representations of the groups $U(n)$ ($SU(n)$) is
realized via a decomposition of a module of the group algebra $\C(S_m)$,
$\boldsymbol{1}=f(\alpha_1)+f(\alpha_2)+\ldots+f(\alpha_s)$, where
$f(\alpha_i)$ are the primitive idempotents of $\C(S_m)$
(Young symmetrizers), $\alpha_i$ are the Young diagrams, ($i=1,\ldots, s$).
The Weyl scheme fully admits the Schur's $\dZ_2$-extension $T_n$,
but an explicit form of the Young's orthogonal representation has been
unanswered so far \cite{Naz90,Naz92}.
\subsection{Decomposition of the spinspace}
The decomposition of the spinspace (\ref{Spin2}) with respect to $SL(2,\C)$
is a simplest case of the Weyl scheme. Every irreducible representation of
the group $SL(2,\C)$ is defined by the Young diagram consisting of only one
row. 
For that reason the representation (\ref{Rep1}) is realized in the space
$\Sym_{(m,0)}$ of all symmetric spintensors of the rank $m$. Dimension of
$\Sym_{(m,0)}$ is equal to $m+1$.

In its turn, every element of the spinspace (\ref{Spin}), related with
the representation (\ref{Rep}), corresponds to an element of
$\dS_{2^k}\otimes\dS_{2^r}$ (representations
$\boldsymbol{\tau}_{\frac{k}{2},0}\otimes\boldsymbol{\tau}_{\frac{r}{2},0}$
and
$\boldsymbol{\tau}_{\frac{k}{2},0}\otimes\boldsymbol{\tau}_{0,\frac{r}{2}}$
are equivalent). This equivalence can be described as follows
\begin{equation}\label{Equiv}
\varphi\otimes\psi\longrightarrow\varphi\otimes\psi I,
\end{equation}
where $\varphi,\psi\in\dS_{2^k}$, $\psi I\in\dS_{2^r}$ and
\[
I=\lambda\begin{pmatrix}
0 & & -1\\
  &\iddots&\\
(-1)_{\frac{r+k}{2}} && 0
\end{pmatrix}
\]
is the matrix of a bilinear form (this matrix is symmetric if
$\frac{r+k}{2}\equiv 0\pmod{2}$ and skewsymmetric if 
$\frac{r+k}{2}\equiv 1\pmod{2}$. In such a way, the representation
(\ref{Rep}) is realized in a symmetric space $\Sym_{(k,r)}$ of
dimension $(k+1)(r+1)$ (or $(2l_1+1)(2l_2+1)$ if suppose $l_1=k/2$ and
$l_2=r/2$). The decomposition of (\ref{Rep}) is given by a Clebsh-Gordan
formula
\[
\boldsymbol{\tau}_{l_1l_1^\prime}\otimes\boldsymbol{\tau}_{l_2l_2^\prime}=
\sum_{|l_1-l_2|\leq k\leq l_1+l_2;|l^\prime_1-l^\prime_2|\leq k^\prime\leq
l^\prime_1+l^\prime_2}\boldsymbol{\tau}_{kk^\prime}.
\]
where the each $\boldsymbol{\tau}_{kk^\prime}$ acts in the space
$\Sym_{(k,k^\prime)}$. In its turn, every space $\Sym_{(k,r)}$
can be represented by a space of polynomials
\begin{gather}
p(z_0,z_1,\bar{z}_0,\bar{z}_1)=\sum_{\substack{(\alpha_1,\ldots,\alpha_k)\\
(\dot{\alpha}_1,\ldots,\dot{\alpha}_r)}}\frac{1}{k!\,r!}
a^{\alpha_1\cdots\alpha_k\dot{\alpha}_1\cdots\dot{\alpha}_r}
z_{\alpha_1}\cdots z_{\alpha_k}\bar{z}_{\dot{\alpha}_1}\cdots
\bar{z}_{\dot{\alpha}_r}\label{SF}\\
(\alpha_i,\dot{\alpha}_i=0,1)\nonumber
\end{gather}
with the basis (the basis of a representation $(l_1l_2)\oplus(l_2l_1)$)
\begin{eqnarray}
z^{l_1l_2}_{m_1m_2}&=&\frac{\zeta^{l_1+m_1}_1\zeta^{l_1-m_1}_2
\zeta^{l_2+m_2}_{\dot{1}}\zeta^{l_2-m_2}_{\dot{2}}
\left(\zeta_1\zeta^\prime_2-\zeta_2\zeta^\prime_1\right)^{2l_1}
\left(\zeta_{\dot{1}}\zeta^\prime_{\dot{2}}-
\zeta_{\dot{2}}\zeta^\prime_{\dot{1}}\right)^{2l_2}}
{\sqrt{(l_1+m_1)!(l_1-m_1)!(l_2+m_2)!(l_2-m_2)!}},\nonumber\\
\bar{z}^{l_2l_1}_{m_2m_1}&=&\frac{\zeta^{l_1+m_1}_{\dot{1}}
\zeta^{l_1-m_1}_{\dot{2}}
\zeta^{l_2+m_2}_{1}\zeta^{l_2-m_2}_{2}
\left(\zeta_{\dot{1}}\zeta^\prime_{\dot{2}}-
\zeta_{\dot{2}}\zeta^\prime_{\dot{1}}\right)^{2l_1}
\left(\zeta_{1}\zeta^\prime_{2}-
\zeta_{2}\zeta^\prime_{1}\right)^{2l_2}}
{\sqrt{(l_1+m_1)!(l_1-m_1)!(l_2+m_2)!(l_2-m_2)!}}.\nonumber
\end{eqnarray}

The vectors $z^{ll^\prime}_{mm^\prime}$ of the canonical helicity basis have the
form
\begin{equation}\label{CG}
z^{ll^\prime}_{mm^\prime}=\sum_{j+k=m,j^\prime+k^\prime=k^\prime}
C(l_1,l_2,l;j,k,m)C(l^\prime_1,l^\prime_2,l^\prime;j^\prime,k^\prime,m^\prime)
\zeta_{jj^\prime}\otimes \zeta_{kk^\prime},
\end{equation}
where
\[
C(l_1,l_2,l;j,k,m)C(l^\prime_1,l^\prime_2,l^\prime;j^\prime,k^\prime,m^\prime)=
B^{j,k,m;j^\prime,k^\prime,m^\prime}_{l_1,l_2,l;l^\prime_1,l^\prime_2,
l^\prime}
\]
are the Clebsch--Gordan coefficients of the group $SL(2,\C)$. Expressing
the Clebsch--Gordan coefficients $C(l_1,l_2,l;j,k,j+k)$ of the group
$SU(2)$ via a generalized hypergeometric function ${}_3F_2$ (see, for example,
\cite{Ros55,Lyu60,VKS75,VK}) 
we see that CG--coefficients of $SL(2,\C)$ have the form
\begin{multline}
B^{j,k,m;j^\prime,k^\prime,m^\prime}_{l_1,l_2,l;l^\prime_1,l^\prime_2,l^\prime}=
(-1)^{l_1+l^\prime_1-j-j^\prime}\times\\
\frac{\Gamma(l_1+l_2-m+1)\Gamma(l^\prime_1+l^\prime_2-m^\prime+1)}
{\Gamma(l_2-l_1+m+1)\Gamma(l^\prime_2-l^\prime_1+m^\prime+1)}\times\\
\left(\frac{(l-m)!(l+l_2-l_1)!(l_1-j)!(l_2+k)!
(l+m)!(2l+1)}{(l_1-l_2+l)!(l_1+l_2-l)!
(l_1+l_2+l)!(l_1-j)!(l_2-k)!}\right)^{\frac{1}{2}}\times\\
\left(\frac{(l^\prime-m^\prime)!(l^\prime+l^\prime_2-l^\prime_1)!
(l^\prime_1-j^\prime)!(l^\prime_2+k^\prime)!
(l^\prime+m^\prime)!(2l^\prime+1)}
{(l^\prime_1-l^\prime_2+l^\prime)!
(l^\prime_1+l^\prime_2-l^\prime)!
(l^\prime_1+l^\prime_2+l^\prime)!
(l^\prime_1-j^\prime)!(l^\prime_2-k^\prime)!}
\right)^{\frac{1}{2}}\times\\
\hypergeom{3}{2}{l+m+1,-l+m,-l_1+j}{-l_1-l_2+m,l_2-l_1+m+1}{1}\times\\
\hypergeom{3}{2}{l^\prime+m^\prime+1,-l^\prime+m^\prime,-l^\prime_1+j^\prime}
{-l^\prime_1-l^\prime_2+m^\prime,l^\prime_2-l^\prime_1+m^\prime+1}{1},
\label{CG'}
\end{multline}
where $m=j+k$, $m^\prime=j^\prime+k^\prime$. 

Infinitesimal operators of $\fG_+$ in the helicity basis have a very simple
form
\begin{eqnarray}
\sA_1&=&-\frac{i}{2}\boldsymbol{\alpha}^l_m\xi_{m-1}-
\frac{i}{2}\boldsymbol{\alpha}^l_{m+1}\xi_{m+1},\nonumber\\
\sA_2&=&\frac{1}{2}\boldsymbol{\alpha}^l_m\xi_{m-1}-
\frac{1}{2}\boldsymbol{\alpha}^l_{m+1}\xi_{m+1},\label{OpA}\\
\sA_3&=&-im\xi_m,\nonumber
\end{eqnarray}
\begin{eqnarray}
\sB_1&=&\frac{1}{2}\boldsymbol{\alpha}^l_m\xi_{m-1}+
\frac{1}{2}\boldsymbol{\alpha}^l_{m+1}\xi_{m+1},\nonumber\\
\sB_2&=&\frac{i}{2}\boldsymbol{\alpha}^l_m\xi_{m-1}-
\frac{i}{2}\boldsymbol{\alpha}^l_{m+1}\xi_{m+1},\label{OpB}\\
\sB_3&=&m\xi_m,\nonumber
\end{eqnarray}
\begin{eqnarray}
\widetilde{A}_1&=&\frac{i}{2}\boldsymbol{\alpha}^l_m\xi_{m-1}+
\frac{i}{2}\boldsymbol{\alpha}^l_{m+1}\xi_{m+1},\nonumber\\
\widetilde{A}_2&=&-\frac{1}{2}\boldsymbol{\alpha}^l_m\xi_{m-1}+
\frac{1}{2}\boldsymbol{\alpha}^l_{m+1}\xi_{m+1},\label{DopA}\\
\widetilde{A}_3&=&im\xi_m,\nonumber\\
\widetilde{B}_1&=&-\frac{1}{2}\boldsymbol{\alpha}^l_m\xi_{m-1}-
\frac{1}{2}\boldsymbol{\alpha}^l_{m+1}\xi_{m+1},\nonumber\\
\widetilde{B}_2&=&-\frac{i}{2}\boldsymbol{\alpha}^l_m\xi_{m-1}+
\frac{i}{2}\boldsymbol{\alpha}^l_{m+1}\xi_{m+1},\label{DopB}\\
\widetilde{B}_3&=&-m\xi_m,\nonumber
\end{eqnarray}
where
\[
\boldsymbol{\alpha}^l_m=\sqrt{(l+m)(l-m+1)}.
\]

\subsection{Gel'fand-Naimark basis}
There exists another representation basis for the Lorentz group
\begin{eqnarray}
H_{3}\xi_{lm} &=&m\xi_{lm},\nonumber\\
H_{+}\xi_{lm} &=&\sqrt{(l+m+1)(l-m)}\xi_{l,m+1},\nonumber\\
H_{-}\xi_{lm} &=&\sqrt{(l+m)(l-m+1)}\xi_{l,m-1},\nonumber\\
F_{3}\xi_{lm} &=&C_{l}\sqrt{l^{2}-m^{2}}\xi_{l-1,m}-A_{l}m\xi_{l,m}-\nonumber \\
&&\hspace{1.8cm}-C_{l+1}\sqrt{(l+1)^{2}-m^{2}}\xi_{l+1,m},\nonumber\\
F_{+}\xi_{lm} &=&C_{l}\sqrt{(l-m)(l-m-1)}\xi_{l-1,m+1}-\nonumber\\
&&\hspace{1.3cm}-A_{l}\sqrt{(l-m)(l+m+1)}\xi_{l,m+1}+\nonumber \\
&&\hspace{1.8cm}+C_{l+1}\sqrt{(l+m+1)(l+m+2)}\xi_{l+1,m+1},\nonumber\\
F_{-}\xi_{lm} &=&-C_{l}\sqrt{(l+m)(l+m-1)}\xi_{l-1,m-1}-\nonumber\\
&&\hspace{1.3cm}-A_{l}\sqrt{(l+m)(l-m+1)}\xi_{l,m-1}-\nonumber\\
&&\hspace{1.8cm}-C_{l+1}\sqrt{(l-m+1)(l-m+2)}\xi_{l+1,m-1},\nonumber
\end{eqnarray}
\[
A_{l}=\frac{il_{0}l_{1}}{l(l+1)},\quad
C_{l}=\frac{i}{l}\sqrt{\frac{(l^{2}-l^{2}_{0})(l^{2}-l^{2}_{1})}
{4l^{2}-1}},
\]
$$m=-l,-l+1,\ldots,l-1,l,$$
$$l=l_{0}\,,l_{0}+1,\ldots,$$
where $l_{0}$ is positive integer or half-integer number, $l_{1}$ is an
arbitrary complex number. 
These formulas 
define a finite--dimensional representation of the group $\fG_+$ when
$l^2_1=(l_0+p)^2$, $p$ is some natural number.
In the case $l^2_1\neq(l_0+p)^2$ we have an
infinite--dimensional representation of $\fG_+$.
The operators $H_{3},H_{+},H_{-},F_{3},F_{+},F_{-}$
are
\begin{eqnarray}
&&H_+=i\sA_1-\sA_2,\quad H_-=i\sA_1+\sA_2,\quad H_3=i\sA_3,\nonumber\\
&&F_+=i\sB_1-\sB_2,\quad F_-=i\sB_1+\sB_2,\quad F_3=i\sB_3.\nonumber
\end{eqnarray}
This basis was firstly given by Gel'fand in 1944 (see also
\cite{Har47,GY48,Nai58}). The following relations between generators
$\sY_\pm$, $\sX_\pm$, $\sY_3$, $\sX_3$ and $H_\pm$, $F_\pm$, $H_3$, $F_3$
define a relationship between the helicity (Van der Waerden) and
Gel'fand-Naimark basises
\[
{\renewcommand{\arraystretch}{1.7}
\begin{array}{ccc}
\sY_+&=&-\dfrac{1}{2}(F_++iH_+),\\
\sY_-&=&-\dfrac{1}{2}(F_-+iH_-),\\
\sY_3&=&-\dfrac{1}{2}(F_3+iH_3),
\end{array}\quad
\begin{array}{ccc}
\sX_+&=&\dfrac{1}{2}(F_+-iH_+),\\
\sX_-&=&\dfrac{1}{2}(F_--iH_-),\\
\sX_3&=&\dfrac{1}{2}(F_3-iH_3).
\end{array}.
}
\]
\subsection{Complexification of $SU(2)$}
As noted previously the explicit form of the isomorphism
$SL(2,\C)\simeq SU(2)\otimes SU(2)$ can be obtained via the
complexification of $SU(2)$. Indeed,
the group $SL(2,\C)$ is a group of all complex matrices
\[\ar
\begin{pmatrix}
\alpha & \beta\\
\gamma & \delta
\end{pmatrix}
\]
of 2-nd order with the determinant $\alpha\delta-\gamma\beta=1$.
The group $SU(2)$ is one of
the real forms of $SL(2,\C)$. The transition from $SU(2)$ to $SL(2,\C)$
is realized via the complexification of three real parameters
$\varphi,\,\theta,\,\psi$ (Euler angles). Let $\theta^c=\theta-i\tau$,
$\varphi^c=\varphi-i\epsilon$, $\psi^c=\psi-i\varepsilon$ be complex
Euler angles, where
\[
{\renewcommand{\arraystretch}{1.05}
\begin{array}{ccccc}
0 &\leq&\re\theta^c=\theta& \leq& \pi,\\
0 &\leq&\re\varphi^c=\varphi& <&2\pi,\\
-2\pi&\leq&\re\psi^c=\psi&<&2\pi,
\end{array}\quad\quad
\begin{array}{ccccc}
-\infty &<&\im\theta^c=\tau&<&+\infty,\\
-\infty&<&\im\varphi^c=\epsilon&<&+\infty,\\
-\infty&<&\im\psi^c=\varepsilon&<&+\infty.
\end{array}}
\]
Replacing in SU(2) the angles $\varphi,\,\theta,\,\psi$ by the
complex angles $\varphi^c,\theta^c,\psi^c$ we come to the following matrix
\begin{gather}
{\renewcommand{\arraystretch}{1.7}
\mathfrak{g}=
\begin{pmatrix}
\cos\dfrac{\theta^c}{2}e^{\frac{i(\varphi^c+\psi^c)}{2}} &
i\sin\dfrac{\theta^c}{2}e^{\frac{i(\varphi^c-\psi^c)}{2}}\\
i\sin\dfrac{\theta^c}{2}e^{\frac{i(\psi^c-\varphi^c)}{2}} &
\cos\dfrac{\theta^c}{2}e^{-\frac{i(\varphi^c+\psi^c)}{2}}
\end{pmatrix}}=
\nonumber\\
{\renewcommand{\arraystretch}{1.9}
\begin{pmatrix}
\s\left[\cos\tfrac{\theta}{2}\ch\tfrac{\tau}{2}+
i\sin\tfrac{\theta}{2}\sh\tfrac{\tau}{2}\right]
e^{\frac{\epsilon+\varepsilon+i(\varphi+\psi)}{2}} &
\s\left[\cos\tfrac{\theta}{2}\sh\tfrac{\tau}{2}+
i\sin\tfrac{\theta}{2}\ch\tfrac{\tau}{2}\right]
e^{\frac{\epsilon-\varepsilon+i(\varphi-\psi)}{2}} \\
\s\left[\cos\tfrac{\theta}{2}\sh\tfrac{\tau}{2}+
i\sin\tfrac{\theta}{2}\ch\tfrac{\tau}{2}\right]
e^{\frac{\varepsilon-\epsilon+i(\psi-\varphi)}{2}} &
\s\left[\cos\tfrac{\theta}{2}\ch\tfrac{\tau}{2}+
i\sin\tfrac{\theta}{2}\sh\tfrac{\tau}{2}\right]
e^{\frac{-\epsilon-\varepsilon-i(\varphi+\psi)}{2}}
\end{pmatrix},}\label{SL1}
\end{gather}
since $\cos\dfrac{1}{2}(\theta-i\tau)=\cos\dfrac{\theta}{2}\ch\dfrac{\tau}{2}+
i\sin\dfrac{\theta}{2}\sh\dfrac{\tau}{2}$, and 
$\sin\dfrac{1}{2}(\theta-i\tau)=\sin\dfrac{\theta}{2}\ch\dfrac{\tau}{2}-
i\cos\dfrac{\theta}{2}\sh\dfrac{\tau}{2}$. It is easy to verify that the
matrix (\ref{SL1}) coincides with a matrix of the fundamental reprsentation
of the group $SL(2,\C)$ (in Euler parametrization):
\begin{multline}
\mathfrak{g}(\varphi,\,\epsilon,\,\theta,\,\tau,\,\psi,\,\varepsilon)=\\[0.2cm]
{\renewcommand{\arraystretch}{1.05}
\begin{pmatrix}
\s e^{i\frac{\varphi}{2}} & \s 0\\
\s 0 &\s e^{-i\frac{\varphi}{2}}
\end{pmatrix}}{\renewcommand{\arraystretch}{1.4}\!\!\!\begin{pmatrix}
\s e^{\frac{\epsilon}{2}} & \s 0\\
\s 0 &\s e^{-\frac{\epsilon}{2}}
\end{pmatrix}}\!\!\!{\renewcommand{\arraystretch}{1.7}\begin{pmatrix}
\s\cos\tfrac{\theta}{2} &\s i\sin\tfrac{\theta}{2}\\
\s i\sin\tfrac{\theta}{2} &\s \cos\tfrac{\theta}{2}
\end{pmatrix}
\!\!\!\begin{pmatrix}
\s\ch\tfrac{\tau}{2} & \s\sh\tfrac{\tau}{2}\\
\s\sh\tfrac{\tau}{2} & \s\ch\tfrac{\tau}{2}
\end{pmatrix}}\!\!\!{\renewcommand{\arraystretch}{1.4}\begin{pmatrix}
\s e^{i\frac{\psi}{2}} & \s 0\\
\s 0 &\s e^{-i\frac{\psi}{2}}
\end{pmatrix}}\!\!\!{\renewcommand{\arraystretch}{1.05}\begin{pmatrix}
\s e^{\frac{\varepsilon}{2}} & \s 0\\
\s 0 &\s e^{-\frac{\varepsilon}{2}}
\end{pmatrix}.}\nonumber
\end{multline}
\begin{sloppypar}
Moreover, the complexification of $SU(2)$ gives us the most simple and
direct way for calculation of matrix elements of the Lorentz group.
It is known that these elements have a great importance in quantum
field theory and widely used at the study of relativistic amplitudes.
The most degenerate representation of such the elements was firstly
obtained by Dolginov, Toptygin and Moskalev \cite{Dol56,DT59,DM59} 
via an analytic
continuation of representations of the group $O_4$. Later on, matrix elements
are studied on the hyperboloid \cite{VS64,VD67}, on the direct product
of the hyperboloid and sphere \cite{LSS68,KMLS69}. The matrix elements
of the principal series are studied by Str\"{o}m \cite{Str65,Str67} in the
Gel'fand-Naimark basis. However, all the matrix elements, calculated in
the GN-basis, have a very complicated form.
\end{sloppypar}

Matrix elements of $SL(2,\C)$ in the helicity basis (see \cite{Var022}) are
\begin{multline}
\fM^l_{mn}(\mathfrak{g})=e^{-m(\epsilon+i\varphi)-n(\varepsilon+i\psi)}
Z^l_{mn}=e^{-m(\epsilon+i\varphi)-n(\varepsilon+i\psi)}\times\\[0.2cm]
\sum^l_{k=-l}i^{m-k}
\sqrt{\Gamma(l-m+1)\Gamma(l+m+1)\Gamma(l-k+1)\Gamma(l+k+1)}\times\\
\cos^{2l}\frac{\theta}{2}\tg^{m-k}\frac{\theta}{2}\times\\[0.2cm]
\sum^{\min(l-m,l+k)}_{j=\max(0,k-m)}
\frac{i^{2j}\tg^{2j}\dfrac{\theta}{2}}
{\Gamma(j+1)\Gamma(l-m-j+1)\Gamma(l+k-j+1)\Gamma(m-k+j+1)}\times\\[0.2cm]
\sqrt{\Gamma(l-n+1)\Gamma(l+n+1)\Gamma(l-k+1)\Gamma(l+k+1)}
\ch^{2l}\frac{\tau}{2}\tnh^{n-k}\frac{\tau}{2}\times\\[0.2cm]
\sum^{\min(l-n,l+k)}_{s=\max(0,k-n)}
\frac{\tnh^{2s}\dfrac{\tau}{2}}
{\Gamma(s+1)\Gamma(l-n-s+1)\Gamma(l+k-s+1)\Gamma(n-k+s+1)}.\label{HS}
\end{multline}
We will call the functions $Z^l_{mn}$ in (\ref{HS}) as
{\it hyperspherical functions}. The
hyperspherical functions $Z^l_{mn}$ can written via the hypergeometric series:
\begin{multline}
Z^l_{mn}=\cos^{2l}\frac{\theta}{2}\ch^{2l}\frac{\tau}{2}
\sum^l_{k=-l}i^{m-k}\tg^{m-k}\frac{\theta}{2}
\tnh^{n-k}\frac{\tau}{2}\times\\[0.2cm]
\hypergeom{2}{1}{m-l+1,1-l-k}{m-k+1}{i^2\tg^2\dfrac{\theta}{2}}
\hypergeom{2}{1}{n-l+1,1-l-k}{n-k+1}{\tnh^2\dfrac{\tau}{2}}.\label{HS1}
\end{multline}
Therefore, relativistic spherical functions can be 
expressed by means of the function
({\it a generalized hyperspherical function})
\begin{equation}\label{HS2}
\fT^l_{mn}(\mathfrak{g})=e^{-m(\epsilon+i\varphi)}Z^l_{mn}
e^{-n(\varepsilon+i\psi)},
\end{equation}
where
\begin{equation}\label{HS3}
Z^l_{mn}=\sum^l_{k=-l}P^l_{mk}(\cos\theta)\mathfrak{P}^l_{kn}(\ch\tau),
\end{equation}
here $P^l_{mn}(\cos\theta)$ is a generalized spherical function on the
group $SU(2)$ (see \cite{GMS}), and $\mathfrak{P}^l_{mn}$ is an analog of
the generalized spherical function for the group $QU(2)$ (so--called
Jacobi function \cite{Vil68}). $QU(2)$ is a group of quasiunitary
unimodular matrices of second order. As well as the group $SU(2)$, the
group $QU(2)$ is one of the real forms of $SL(2,\C)$
($QU(2)$ is noncompact).
Further, from (\ref{HS1}) we see that the function $Z^l_{mn}$ depends on
two variables $\theta$ and $\tau$. Therefore, using Bateman factorization we can
express the hyperspherical functions $Z^l_{mn}$ via Appell functions
$F_1$--$F_4$ (hypergeometric series of two variables \cite{AK26,Bat}).
\section{Gel'fand-Yaglom equations in the helicity basis}
As a direct consequence of the isomorphism $SL(2,\C)\simeq SU(2)\otimes SU(2)$
we have equations
\begin{eqnarray}
\sum^3_{i=1}\sL_i\frac{\partial\boldsymbol{\psi}}{\partial x_i}-
i\sum^3_{i=1}\sL_i\frac{\partial\boldsymbol{\psi}}{\partial x^\ast_i}
+\kappa^c\boldsymbol{\psi}&=&0,\nonumber\\
\sum^3_{i=1}\sL^\ast_i\frac{\partial\dot{\boldsymbol{\psi}}}
{\partial\widetilde{x}_i}+
i\sum^3_{i=1}\sL^\ast_i\frac{\partial\dot{\boldsymbol{\psi}}}
{\partial\widetilde{x}^\ast_i}+
\dot{\kappa}^c\dot{\boldsymbol{\psi}}&=&0,\label{Complex}
\end{eqnarray}
with invariance conditions
\begin{eqnarray}
\sum_k g^-_{ik}T_{\fg}\sL_kT^{-1}_{\fg}&=&\sL_i,\nonumber\\
\sum_k g^+_{ik}\overset{\ast}{T}_{\fg}\sL^\ast_k
\overset{\ast}{T}_{\fg}\!\!\!{}^{-1}&=&\sL^\ast_i.\label{IC}
\end{eqnarray}
The equations (\ref{Complex}) act in a 3-dimensional complex space $\C^3$.
In its turn, the space $\C^3$ is isometric to a 6-dimensional bivector
space $\R^6$ (parameter space or group manifold of the Lorentz group
\cite{Kag26,Pet69}).

Let us find commutation relations between the matrices $\sL_i$, $\sL^\ast_i$
and infinitesimal operators (\ref{OpA}), (\ref{OpB}), (\ref{DopA}),
(\ref{DopB}). First of all, let us present transformations $T_{\fg}$
($\overset{\ast}{T}_{\fg}$) in the infinitesimal form,
$\sI+\sA_i\xi+\ldots$, $\sI+\sB_i\xi+\ldots$, $\sI+\widetilde{\sA}_i\xi+
\ldots$, $\sI+\widetilde{\sB}_i+\ldots$. Substituting these transformations
into invariance conditions (\ref{IC}) we obtain with an accuracy of the
terms of second order the following commutation relations
\begin{equation}\label{AL}
\begin{array}{rcl}
\ld\sA_1,\sL_1\rd &=& 0,\\
\ld\sA_2,\sL_1\rd &=&-\sL_3,\\
\ld\sA_3,\sL_1\rd &=& \sL_2,
\end{array}\;\;\;
\begin{array}{rcl}
\ld\sA_1,\sL_2\rd &=& \sL_3,\\
\ld\sA_2,\sL_2\rd &=& 0,\\
\ld\sA_3,\sL_2\rd &=&-\sL_1,
\end{array}\;\;\;
\begin{array}{rcl}
\ld\sA_1,\sL_3\rd &=&-\sL_2,\\
\ld\sA_2,\sL_3\rd &=& \sL_1,\\
\ld\sA_3,\sL_3\rd &=& 0.
\end{array}
\end{equation}
\begin{equation}\label{BL}
\begin{array}{rcl}
\ld\sB_1,\sL_1\rd &=& 0,\\
\ld\sB_2,\sL_1\rd &=&-i\sL_3,\\
\ld\sB_3,\sL_1\rd &=& i\sL_2,
\end{array}\;\;\;
\begin{array}{rcl}
\ld\sB_1,\sL_2\rd &=& i\sL_3,\\
\ld\sB_2,\sL_2\rd &=& 0,\\
\ld\sB_3,\sL_2\rd &=&-i\sL_1,
\end{array}\;\;\;
\begin{array}{rcl}
\ld\sB_1,\sL_3\rd &=&-i\sL_2,\\
\ld\sB_2,\sL_3\rd &=& i\sL_1,\\
\ld\sB_3,\sL_3\rd &=& 0.
\end{array}
\end{equation}
\begin{equation}\label{DAL}
{\renewcommand{\arraystretch}{1.6}
\begin{array}{rcl}
\ld\widetilde{\sA}_1,\sL^\ast_1\rd &=& 0,\\
\ld\widetilde{\sA}_2,\sL^\ast_1\rd &=& \sL^\ast_3,\\
\ld\widetilde{\sA}_3,\sL^\ast_1\rd &=&-\sL^\ast_2,
\end{array}\;\;\;
\begin{array}{rcl}
\ld\widetilde{\sA}_1,\sL^\ast_2\rd &=&-\sL^\ast_3,\\
\ld\widetilde{\sA}_2,\sL^\ast_2\rd &=& 0,\\
\ld\widetilde{\sA}_3,\sL^\ast_2\rd &=& \sL^\ast_1,
\end{array}\;\;\;
\begin{array}{rcl}
\ld\widetilde{\sA}_1,\sL^\ast_3\rd &=& \sL^\ast_2,\\
\ld\widetilde{\sA}_2,\sL^\ast_3\rd &=&-\sL^\ast_1,\\
\ld\widetilde{\sA}_3,\sL^\ast_3\rd &=& 0.
\end{array}}
\end{equation}
\begin{equation}\label{DBL}
{\renewcommand{\arraystretch}{1.6}
\begin{array}{rcl}
\ld\widetilde{\sB}_1,\sL^\ast_1\rd &=& 0,\\
\ld\widetilde{\sB}_2,\sL^\ast_1\rd &=&-i\sL^\ast_3,\\
\ld\widetilde{\sB}_3,\sL^\ast_1\rd &=&+i\sL^\ast_2,
\end{array}\;\;\;
\begin{array}{rcl}
\ld\widetilde{\sB}_1,\sL^\ast_2\rd &=&+i\sL^\ast_3,\\
\ld\widetilde{\sB}_2,\sL^\ast_2\rd &=& 0,\\
\ld\widetilde{\sB}_3,\sL^\ast_2\rd &=&-i\sL^\ast_1,
\end{array}\;\;\;
\begin{array}{rcl}
\ld\widetilde{\sB}_1,\sL^\ast_3\rd &=&-i\sL^\ast_2,\\
\ld\widetilde{\sB}_2,\sL^\ast_3\rd &=&+i\sL^\ast_1,\\
\ld\widetilde{\sB}_3,\sL^\ast_3\rd &=& 0.
\end{array}}
\end{equation} 
\begin{sloppypar}
Further, using the latter relations and taking into account (\ref{SL25})
it is easy to establish commutation relations between $\sL_3$ and
generators $\sY_\pm$, $\sY_3$, $\sX_\pm$, $\sX_3$:
\end{sloppypar}
\begin{equation}\label{LY}
{\renewcommand{\arraystretch}{1.5}
\left.\begin{array}{l}
\ld\sY_+,\ld\sL_3,\sY_-\rd\rd =2\sL_3,\\
\ld\sL_3,\sY_3\rd =0,\\
\ld\sL_3,\sX_-\rd =0,\\
\ld\sL_3,\sX_+\rd =0,\\
\ld\sL_3,\sX_3\rd =0,
\end{array}\right\}
}
\end{equation}
\begin{equation}\label{LX}
{\renewcommand{\arraystretch}{1.5}
\left.\begin{array}{l}
\ld\sX_+,\ld\sL^\ast_3,\sX_-\rd\rd = 2\sL^\ast_3,\\
\ld\sL^\ast_3,\sX_3\rd =0,\\
\ld\sL^\ast_3,\sY_-\rd =0,\\
\ld\sL^\ast_3,\sY_+\rd =0,\\
\ld\sL^\ast_3,\sY_3\rd =0.
\end{array}\right\}
}
\end{equation}
From (\ref{LY}) and (\ref{LX}) it immediately follows that elements of
the matrices $\sL_3$ and $\sL^\ast_3$ are
\begin{equation}\label{L3}
{\renewcommand{\arraystretch}{1.6}
\sL_3:\quad\left\{\begin{array}{ccc}
c^{k^\prime k}_{l-1,l,m}&=&
c^{k^\prime k}_{l-1,l}\sqrt{l^2-m^2},\\
c^{k^\prime k}_{l,l,m}&=&c^{k^\prime k}_{ll}m,\\
c^{k^\prime k}_{l+1,l,m}&=&
c^{k^\prime k}_{l+1,l}\sqrt{(l+1)^2-m^2}.
\end{array}\right.}
\end{equation}
\begin{equation}\label{L3'}
{\renewcommand{\arraystretch}{1.6}
\sL^\ast_3:\quad\left\{\begin{array}{ccc}
f^{\dot{k}^\prime\dot{k}}_{\dot{l}-1,\dot{l},\dot{m}}&=&
c^{\dot{k}^\prime\dot{k}}_{\dot{l}-1,\dot{l}}
\sqrt{\dot{l}^2-\dot{m}^2},\\
f^{\dot{k}^\prime\dot{k}}_{\dot{l}\dot{l},\dot{m}}&=&
c^{\dot{k}^\prime\dot{k}}_{\dot{l}\dot{l}}\dot{m},\\
f^{\dot{k}^\prime\dot{k}}_{\dot{l}+1,\dot{l},\dot{m}}&=&
c^{\dot{k}^\prime\dot{k}}_{\dot{l}+1,\dot{l}}
\sqrt{(\dot{l}+1)^2-\dot{m}^2}.
\end{array}\right.}
\end{equation}
All other elements are equal to zero. If we know the elements of $\sL_3$
and $\sL^\ast_3$, then the elements of $\sL_1$, $\sL_2$ and
$\sL^\ast_1$, $\sL^\ast_2$ can be obtained via the relations 
(\ref{AL})--(\ref{DBL}).

With a view to separate the variables in (\ref{Complex}) let us assume
\begin{equation}\label{F}
\psi^k_{lm}=\boldsymbol{f}^{l_0}_{lmk}(r)
\fM^{l_0}_{mn}(\varphi,\epsilon,\theta,\tau,0,0),
\end{equation}
where $l_0\ge l$, and $-l_0\le m$, $n\le l_0$.
The full separation of variables in (\ref{Complex}) has been given in
the recent paper \cite{Var024}, where the hyperspherical functions
$\fM^l_{m,n}$ are defined on a 2-dimensional complex sphere
(about the 2-dimensional complex sphere see \cite{SH70}). In the result of
the separation we come to a system of ordinary differential equations which
depends on the radial functions,
\begin{multline}
\sum_{k^\prime}c^{kk^\prime}_{l,l-1}\left[
2\sqrt{l^2-m^2}
\frac{d\boldsymbol{f}^{l_0}_{l-1,m,k^\prime}(r)}{d r}+\right.\\
-\frac{1}{r}(l+1)\sqrt{l^2-m^2}\boldsymbol{f}^{l_0}_{l-1,m,k^\prime}(r)+\\
+\frac{i}{r}\sqrt{(l+m)(l+m-1)}
\sqrt{(\dot{l}_0+\dot{m})(\dot{l}_0-\dot{m}+1)}
\boldsymbol{f}^{l_0}_{l-1,m-1,k^\prime}(r)+\\
\left.+\frac{i}{r}\sqrt{(l-m)(l-m-1)}
\sqrt{(\dot{l}_0+\dot{m}+1)(\dot{l}_0-\dot{m})}
\boldsymbol{f}^{l_0}_{l-1,m+1,k^\prime}(r)\right]+
\nonumber
\end{multline}
\begin{multline}
\sum_{k^\prime}c^{kk^\prime}_{ll}\left[
2m\frac{d\boldsymbol{f}^{l_0}_{l,m,k^\prime}(r)}{d r}-
\frac{1}{r}m\boldsymbol{f}^{l_0}_{l,m,k^\prime}(r)-\right.
\\
-\frac{i}{r}\sqrt{(l+m)(l-m+1)}
\sqrt{(\dot{l}_0+\dot{m})(\dot{l}_0-\dot{m}+1)}
\boldsymbol{f}^{l_0}_{l,m-1,k^\prime}(r)+\\
\left.+\frac{i}{r}\sqrt{(l+m+1)(l-m)}
\sqrt{(\dot{l}_0+\dot{m}+1)(\dot{l}_0-\dot{m})}
\boldsymbol{f}^{l_0}_{l,m+1,k^\prime}(r)\right]+
\nonumber
\end{multline}
\begin{multline}
\sum_{k^\prime}c^{kk^\prime}_{l,l+1}\left[
2\sqrt{(l+1)^2-m^2}
\frac{d\boldsymbol{f}^{l_0}_{l+1,m,k^\prime}(r)}{d r}-\right.
\\
+\frac{1}{r}l\sqrt{(l+1)^2-m^2}
\boldsymbol{f}^{l_0}_{l+1,m,k^\prime}(r)-\\
-\frac{i}{r}\sqrt{(l-m+1)(l-m+2)}
\sqrt{(\dot{l}_0+\dot{m})(\dot{l}_0-\dot{m}+1)}
\boldsymbol{f}^{l_0}_{l+1,m-1,k^\prime}(r)-\\
\left.-\frac{i}{r}\sqrt{(l+m+1)(l+m+2)}
\sqrt{(\dot{l}_0+\dot{m}+1)(\dot{l}_0-\dot{m})}
\boldsymbol{f}^{l_0}_{l+1,m+1,k^\prime}(r)\right]+
\\
+\kappa^c\boldsymbol{f}^{l_0}_{lmk}(r)=0,\nonumber
\end{multline}
\begin{multline}
\sum_{\dot{k}^\prime}c^{\dot{k}\dot{k}^\prime}_{\dot{l},\dot{l}-1}\left[
2\sqrt{\dot{l}^2-\dot{m}^2}
\frac{d\boldsymbol{f}^{\dot{l}_0}_{\dot{l}-1,\dot{m},\dot{k}^\prime}(r)}
{d r^\ast}-\right.\\
-\frac{1}{r^\ast}(\dot{l}+1)\sqrt{\dot{l}^2-\dot{m}^2}
\boldsymbol{f}^{\dot{l}_0}_{\dot{l}-1,\dot{m},\dot{k}^\prime}(r)+\\
+\frac{i}{r^\ast}\sqrt{(\dot{l}+\dot{m})(\dot{l}+\dot{m}-1)}
\sqrt{(l_0+m)(l_0-m+1)}
\boldsymbol{f}^{\dot{l}_0}_{\dot{l}-1,\dot{m}-1,\dot{k}^\prime}(r)+\\
\left.
+\frac{i}{r^\ast}\sqrt{(\dot{l}-\dot{m})(\dot{l}-\dot{m}-1)}
\sqrt{(l_0+m+1)(l_0-m)}
\boldsymbol{f}^{\dot{l}_0}_{\dot{l}-1,\dot{m}+1,\dot{k}^\prime}(r)\right]+
\nonumber
\end{multline}
\begin{multline}
\sum_{\dot{k}^\prime}c^{\dot{k}\dot{k}^\prime}_{\dot{l},\dot{l}}\left[
2\dot{m}
\frac{d\boldsymbol{f}^{\dot{l}_0}_{\dot{l},\dot{m},\dot{k}^\prime}(r)}
{d r^\ast}
-\frac{1}{r^\ast}\dot{m}
\boldsymbol{f}^{\dot{l}_0}_{\dot{l},\dot{m},\dot{k}^\prime}(r)-\right.
\\
-\frac{i}{r^\ast}\sqrt{(\dot{l}+\dot{m})(\dot{l}-\dot{m}-1)}
\sqrt{(l_0+m)(l_0-m+1)}
\boldsymbol{f}^{\dot{l}_0}_{\dot{l},\dot{m}-1,\dot{k}^\prime}(r)+\\
\left.
+\frac{i}{r^\ast}\sqrt{(\dot{l}+\dot{m}+1)(\dot{l}-\dot{m})}
\sqrt{(l_0+m+1)(l_0-m)}
\boldsymbol{f}^{\dot{l}_0}_{\dot{l},\dot{m}+1,\dot{k}^\prime}(r)\right]+
\nonumber
\end{multline}
\begin{multline}
\sum_{\dot{k}^\prime}c^{\dot{k}\dot{k}^\prime}_{\dot{l},\dot{l}+1}\left[
2\sqrt{(\dot{l}+1)^2-\dot{m}^2}
\frac{d\boldsymbol{f}^{\dot{l}_0}_{\dot{l}+1,\dot{m},\dot{k}^\prime}(r)}
{d r^\ast}+\right.
\\
+\frac{1}{r^\ast}\dot{l}\sqrt{(\dot{l}+1)^2-\dot{m}^2}
\boldsymbol{f}^{\dot{l}_0}_{\dot{l}+1,\dot{m},\dot{k}^\prime}(r)-\\
-\frac{i}{r^\ast}\sqrt{(\dot{l}-\dot{m}+1)(\dot{l}-\dot{m}+2)}
\sqrt{(l_0+m)(l_0-m+1)}
\boldsymbol{f}^{\dot{l}_0}_{\dot{l}+1,\dot{m}-1,\dot{k}^\prime}(r)-\\
\left.
-\frac{i}{r^\ast}\sqrt{(\dot{l}+\dot{m}+1)(\dot{l}+\dot{m}+2)}
\sqrt{(l_0+m+1)(l_0-m)}
\boldsymbol{f}^{\dot{l}_0}_{\dot{l}+1,\dot{m}+1,\dot{k}^\prime}(r)\right]+
\\
+\dot{\kappa}^c
\boldsymbol{f}^{\dot{l}_0}_{\dot{l}\dot{m}\dot{k}^\prime}(r)=0.
\label{RFS}
\end{multline}
This system is solvable for any spin in a class of the Bessel functions.
Therefore, according to (\ref{F}) relativistic wave functions are
expressed via the products of cylindrical and hyperspherical functions.
For that reason
the wave function hardly depends on the parameters of the Lorentz group.
Moreover, it allows to consider solutions of RWE as the functions on
the Lorentz group. On the other hand, a
group theoretical description of RWE allows to present all the physical
fields on an equal footing. Namely, all these fields are the functions
on the Lorentz (Poincar\'{e}) group. Such a description corresponds
to a quantum field theory on the Poincar\'{e} group introduced by
Lur\c{c}at \cite{Lur64} (see also \cite{BK69,Kih70,Tol96,GS01} and 
references therein).
\subsection{Decomposable and indecomposable RWE}
In accordance with equivalence (\ref{Equiv}) and decompositions of
(\ref{Spin}) and (\ref{Rep}) the matrix $\sL_3$ can be written in the
form
\begin{equation}\label{Block}
\sL_3=\begin{pmatrix}
C^1 & & & &\\
 & C^2& &\mbox{\LARGE 0}&\\
 & &\ddots& &\\
 &\mbox{\LARGE 0}& &C^s&\\
 & & & &\ddots
\end{pmatrix}.
\end{equation}
The matrix $\sL^\ast_3$ has the same form. $C^s$ is a spin block.
The each spin block $C^s$ is
realized in the space $\Sym_s$. The matrix elements of the
corresponded representations are expressed via the hyperspherical functions.
If the spin block $C^s$ has non--null
roots, then the particle possesses the spin $s$ \cite{GY48,AD72,PS83}. 
The spin block
$C^s$ in (\ref{Block}) consists of the elements
$c^s_{\boldsymbol{\tau}\boldsymbol{\tau}^\prime}$, where
$\boldsymbol{\tau}_{l_1,l_2}$ and $\boldsymbol{\tau}_{l^\prime_1,l^\prime_2}$
are interlocking irreducible representations of the Lorentz group,
that is, such representations, for which
$l^\prime_1=l_1\pm\frac{1}{2}$, $l^\prime_2=l_2\pm\frac{1}{2}$.
At this point the block $C^s$ contains only the elements
$c^s_{\boldsymbol{\tau}\boldsymbol{\tau}^\prime}$ corresponding to such
interlocking representations $\boldsymbol{\tau}_{l_1,l_2}$,
$\boldsymbol{\tau}_{l^\prime_1,l^\prime_2}$ which satisfy the conditions
\[
|l_1-l_2|\leq s\leq l_1+l_2,\quad
|l^\prime_1-l^\prime_2|\leq s \leq l^\prime_1+l^\prime_2.
\]

According to a de Broglie theory of fusion \cite{Bro43,Soc65}
interlocking representations give rise to indecomposable RWE.
Otherwise, we have decomposable equations. As known, the indecomposable
RWE correspond to composite particles. A wide variety of such
representations and RWE it seems to be sufficient for description of
all particle world.

\end{document}